\documentclass[
  ,final            % use final for the camera ready runs
  ]
  {aipproc}

\layoutstyle{6x9}

\begin{document}
\begin{flushright}
 IFT-UAM/CSIC-12-17
\end{flushright}
 
\title{The Sound of Strongly Coupled Field Theories: Quasinormal Modes in AdS}

\author{Karl Landsteiner}{
  address={Instituto de F\'\i sica Te\'orica UAM-CSIC\\ Universidad Aut\'onoma de Madrid\\
c/ Nicolas Cabrera 13-15\\ 28049 Cantoblanco, Spain}
}

% \author{<author2>}{
%   address={<common address for author2 and author3>}
% }
% 
% \author{<author3>}{
%   address={<common address for author2 and author3>}
%   ,altaddress={<author1 address>} % additional visiting address
% }

\begin{abstract}
The AdS/CFT correspondence has developed over the last years into a very useful and powerful
tool for studying strongly coupled field theories at finite temperature and density. Of particular interest
is the regime of near equilibrium real time evolution that can be captured via linear response theory.
The AdS/CFT correspondence allows the calculation of retarded two point functions of gauge invariant operators by studying fluctuations around asymptotically AdS black holes. A major role is played by the 
poles of these holographic response functions: the quasinormal frequencies. I will review the applications of these ideas to the hydrodynamics of the strongly coupled quark gluon plasma and the holographic realization of strongly coupled superfluids.
\end{abstract}

\maketitle

%%%%%%%%%%%%%%%%%%%%%%%%%%%%%%%%%%%%%%%%%%%%
%% MAINMATTER
%%%%%%%%%%%%%%%%%%%%%%%%%%%%%%%%%%%%%%%%%%%%

\section{Introduction}\label{sec:intro}

String theory is most often cited as the best candidate for the big TOE, the all encompassing theory of everything. Although there are many physicists who are quite outspoken sceptics of that claim, I think nobody can refute the fact that string theory is a veritable theory of everything in a slightly different sense: string theory unites and uses all theoretical concepts that are normally specific to certain sub-fields of physics, such as elementary particle physics, condensed matter physics and gravitational
physics.

One of these unifications so characteristic of string theory is the application of the AdS/CFT correspondence \cite{Aharony:1999ti} to the realm of heavy ion collisions and to condensed matter physics. This sounds like a very exotic mixture of topics to talk about
at a gravity meeting as was the ERE 2011. But the tools that are necessary to undertake a journey into the realm of heavy ion and condensed matter physics are indeed very well known to specialists working in general relativity. 
I will concentrate in my exposition on one particular of these tools: quasinormal modes of asymptotically anti-de Sitter black holes. 

Quasinormal modes arise because of the most basic properties of black holes: they swallow up everything
leaving no traces back except for conserved charges. 
If we perturb a static asymptotically flat
black hole space-time by any excitation, a scalar field, gauge field or metric fluctuation, these fluctuations can only end up
being either swallowed by the black hole or radiated off to infinite. Quasinormal modes are the spectrum of linearized
fluctuations with outgoing boundary conditions at spatial infinity and infalling boundary conditions on the black hole horizon. 
The mathematical problem one has to solve is one dearest to the hart of every theoretical physicist: an eigenvalue problem for a differential operator. However, contrary to the eigenvalue problems one encounters normally, the particular
boundary conditions make the the differential operator a non-hermitian one and therefore the eigen frequencies turn out to be
complex numbers of the form
\begin{equation}\label{eq:qnm}
 \omega_n = \Omega_n - i \Gamma_n\,.
\end{equation}
The real part is the frequency with which the perturbation oscillates whereas the imaginary part is the rate at which this oscillation is damped. 
The fluctuations are damped if $\Gamma_n > 0$, otherwise an instability in form of an exponential growth in time appears. In asymptotically AdS spaces spacelike infinity acts like a reflecting boundary and Neumann, Dirichlet
or some form of mixed (Robin) boundary conditions have to by supplied there. Quasinormal modes for asymptotically AdS black holes have first been considered in \cite{Chan:1996yk} and in the context of the AdS/CFT correspondence in \cite{Horowitz:1999jd,Cardoso:2001hn}. 
The holographic dictionary singles out a specific boundary condition corresponding to the poles of the holographic Green's functions \cite{Birmingham:2001pj,Starinets:2002br}.

\begin{figure}\label{fig:bcs}%
\includegraphics[height=6cm]{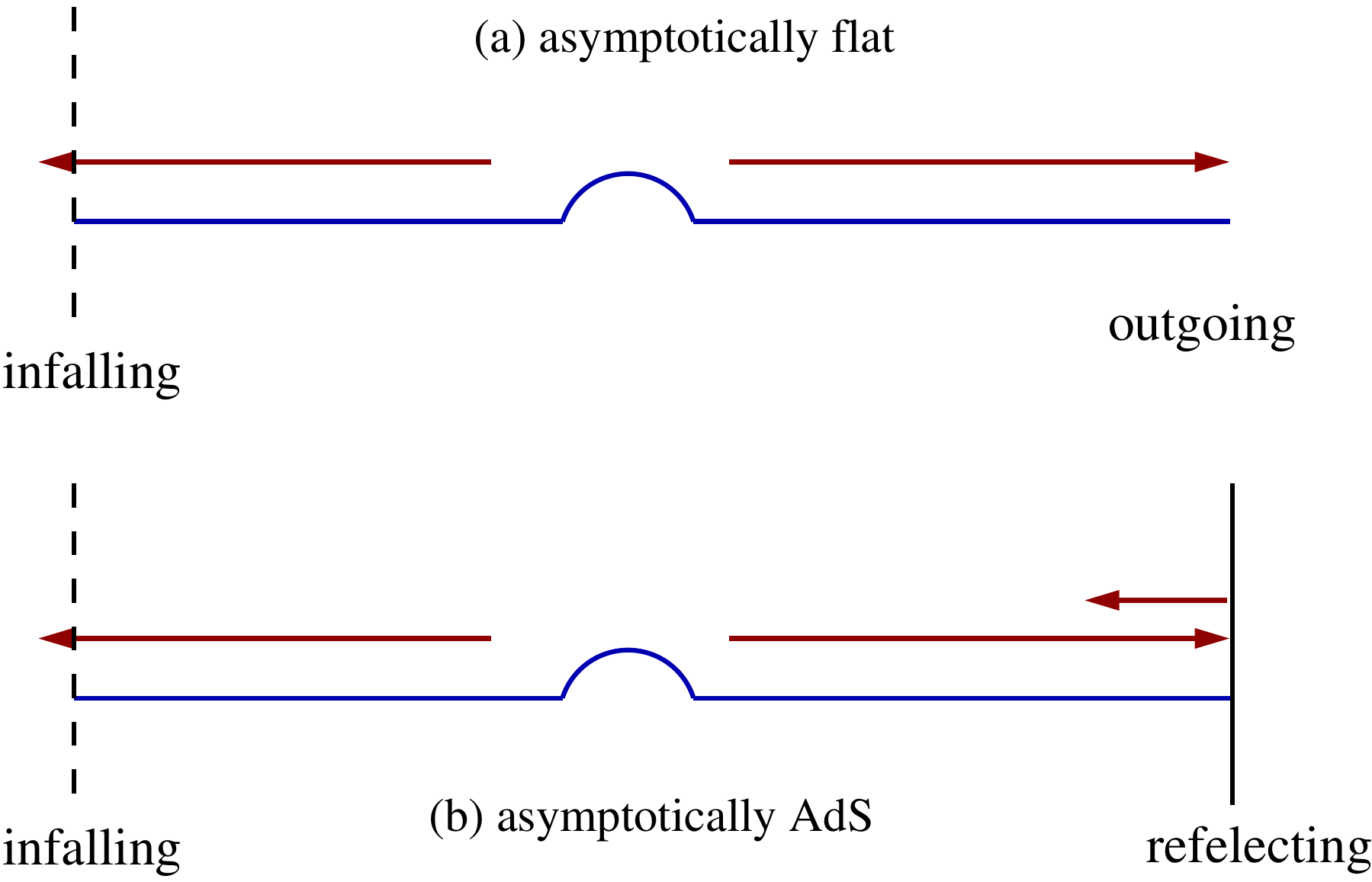}%
\caption{Quasinormal modes are small fluctuations obeying infalling boundary conditions on the horizon. (a) In asymptotically
flat space times they also obey outgoing boundary conditions at infinity. (b) In asymptotically AdS space times the boundary conditions at $r\rightarrow \infty$ can be either Dirichlet, Neuman or mixed ones. In the context of Holography the they are fixed  to be the poles of the holographic retarded correlators}%
\end{figure}%

Unfortunately the mathematical theory of quasinormal modes lacks far behind its hermitian cousin. Few general
statements about the spectrum (\ref{eq:qnm}) can therefore be made, e.g. only for some simple equations stability can be proved.
The quasinormal mode spectrum in most cases has to be calculated using some numerical approximation method. For the AdS/CFT applications it is however enough to study the properties of some of the lowest lying quasinormal modes. It turns out that these are intimately connected to the hydrodynamic behavior of strongly coupled field theories.
Very useful reviews on quasinormal modes are \cite{Nollert:1999ji,Berti:2009kk}

\section{The AdS/CFT correspondence}\label{sec:adscft}

Before discussing the role of quasinormal modes we need first to understand the basics of the AdS/CFT correspondence.
We will be very brief in summarizing the main ideas.  
The fundamental statement is ``Type IIB string theory on AdS$_5$ $\times$ S$_5$ is dual to ${\cal N}=4$ supersymmetric gauge theory''. Let's try to understand this statement in a crash-course style. The ${\cal N}=4$ supersymmetric gauge theory is 
a non-abelian, four dimensional quantum field theory whose field content consists of six scalars, four Majorana fermions and a
vector field. They all transform under the adjoint representation of the gauge group $SU(N)$.  It features four supersymmetries and this fixes all the couplings between the different fields. As it is a gauge theory physical observables are gauge invariant
operators such as $\mathrm{tr}(F_{\mu\nu} F^{\mu\nu})$. The global symmetry group $SO(6)$ acts on the scalars and the
fermions (in the $SU(4)$ spin representation of $SO(6)$). Because of the high amount of supersymmetries the theory also has
exact conformal symmetry on the quantum level and this leads to four additional conformal supersymmetries , such that we have $2\times 4\times 4 = 32$ supercharges in total. 

The dual theory is a theory of gravity (this is what type IIB string theory is) but living in quite a few more dimensions, ten as
opposed to the four the field theory knows about. But five of these ten are easily got rid off: the isometries of the $S_5$ part
of the geometry form $SO(6)$. The $S_5$ is the geometric realization of what appears as an internal, global symmetry group in the field theory. Type IIB string theory has two supersymmetries in ten dimensions, they are Majorana-Weyl spinors of same chirality with 16 components each, which gives the 32 supercharges necessary to match the field theory counting. 

The field theory is characterized by two parameters, the gauge coupling $g_{YM}$ and the rank of the gauge group $N$. The dual 
string theory has a string coupling $g_s$ (the amplitude for a string to split in two) and a fundamental length scale $l_s$, the string scale. Furthermore the geometry is determined by a scale $L$ determining the curvature of the AdS$_5$ as $R=-20/L^2$. The AdS/CFT correspondence relates these parameters
in the following way:
\begin{eqnarray} \label{eq:holodictionary}
g^2_{YM} N &\propto& \frac{L^4}{l_s^4}\\
1/N &\propto& g_s
\end{eqnarray}
The AdS/CFT correspondence is therefore a strong weak coupling duality: for weak curvature we have large $L$ and therefore also
large 't-Hooft coupling. In this regime of weak curvature stringy effects are negligible and we can approximate the string theory
by type IIB supergravity. If we furthermore take the rank of the gauge group $N$ to be very large we can also neglect quantum
loop effects and end up with classical supergravity! This is the form of the correspondence most useful for the applications 
to many body physics: classical (super)gravity on $(d+1)$ dimensional Anti-de Sitter space is the infinite coupling and infinite rank limit of a gauge theory in $d$ dimensions. Here we have allowed ourselves to be already a bit more general. Once we have understood the original example based on the maximally supersymmetric four dimensional field theory we can conjecture that every theory of gravity plus some suitably chosen matter fields on $(d+1)$ dimensional Anti-de Sitter space is a dual to a certain quantum field theory in $d$ dimensions. In fact we might even be a bit more brave and delete the words ``dual to a'' in the previous phrase. This is the point of view taken in the applications of the AdS/CFT correspondence to the world of solid state physics. The additional matter fields are then chosen to reflect a particular symmetry content of the underlying quantum field theoretical system one is interested in. 
Having this in mind we will forget from now on some of the non-essential ingredients, such as supersymmetry and the extra dimensions in form of the $S_5$.

For the applications to quantum field theory the most useful foliation of Anti-de Sitter space is the so called Poincar\'e patch
in which the line element takes the form
\begin{equation}\label{eq:ads}
 ds^2 = \frac{r^2}{L^2} ( -dt^2 + d\vec{x}^2 ) + \frac{L^2}{r^2}dr^2\,.
\end{equation}
The space on which the dual quantum field theory lives is recovered by taking the limit 
$ds^2_\mathrm{QFT} = \lim_{r\rightarrow \infty} r^{-2} ds^2$. This is why sometimes it is said that the dual field theory
lives on the ``boundary'' of AdS and why the AdS/CFT correspondence is also referred to as ``Holography``. The radial coordinate
has however a physical interpretation as an energy scale. We can identify the high-energy UV limit in the field theory with the
$r\rightarrow \infty$ limit in the AdS geometry, whereas the low-energy IR limit is $r\rightarrow 0$. 

The asymptotic behavior of the fields in AdS has the form 
\begin{equation}\label{eq:asymptoticexpansio}
 \Phi = r^{-\Delta_-} \left(\Phi_0(x) + O(r^{-2})\right) + r^{-\Delta_+} \left(\Phi_1(x) + O(r^{-2} )\right)\,.
\end{equation}
The exponents $\Delta_{\pm}$ obey $\Delta_- < \Delta_+$ and depend on the nature of the field, e.g. for a scalar field of 
mass $m$ they are $\Delta_{\pm} = \frac 1 2 (d\pm \sqrt{d^2+4m^2L^2})$. 

We now would have to evaluate the path integral over
the fields in AdS keeping the boundary values $\Phi_0(x)$ fixed. The result is a functional depending on the boundary
fields $\Phi_0(x)$. Now the boundary field $\Phi_0(x)$ is interpreted as the source $J(x)$ that couples to a (gauge invariant)
operator ${\cal O}(x)$ of conformal dimension $\Delta_+$ in the field theory
\begin{equation}
 Z[J] = \int_{\Phi_0 = J} d \Phi\; \exp(-i S[\Phi])
\end{equation}
Connected Green's functions of gauge invariant operators in the quantum field theory can now be generated by functional
differentiation with respect to the sources
\begin{equation}
\left\langle {\cal O}_1(x_1) \dots  {\cal O}_n(x_n) \right\rangle = \frac{\delta^n \log Z}{\delta J_1(x_1) \dots \delta J_n(x_n)}\,.
\end{equation}
In the limit in which the gravity theory becomes classical, i.e. the large $N$ and large coupling $g^2_{YM} N$ limit the
path integral is dominated by the classical solutions to the field equations and $\log Z$ can be replaced by the
classical action evaluated on a solution of the field equations. In this case the coefficient $\Phi_1(x)$ of the asymptotic
expansion (\ref{eq:asymptoticexpansio}) is the vacuum expectation value of the operator sourced by $\Phi_0$.
\begin{equation}
 \left\langle {\cal O}(x)\right\rangle \propto \Phi_1(x)\,.
\end{equation}

To explicitly compute $\Phi_1(x)$ we need to supply a second boundary condition, so far we have fixed only the asymptotic value
$\Phi_0$. For time independent solutions we demand regularity in the interior of the (possibly only asymptotically) AdS space.
For time dependent solutions we need a different boundary condition. The metric (\ref{eq:ads}) has a (degenerate) 
horizon at $r=0$ and it was argued in \cite{Son:2002sd} that for time dependent solutions the appropriate boundary
conditions are infalling ones. In particular it was argued that the Green-functions that are calculated with these infalling
boundary conditions correspond to retarded ones in the field theory, e.g. on the level of two point functions the retarded
correlator is given by
\begin{equation}\label{eq:retarded}
 G_r(t,\vec{x}) = -i \Theta(t)\left \langle [ {\cal O}(t,\vec{x}), {\cal O}(0,0) ] \right \rangle\,.
\end{equation}

That does not yet lead to quasinormal modes as long as we are in pure AdS space. But that
changes as soon as we consider a black hole with planar horizon topology, i.e. a black brane in AdS. Its line element is
\begin{eqnarray}\label{eq:blackbrane}
 ds^2/L^2 &=& r^2\left( - f(r)dt^2 + d\vec{x}^2\right) + \frac{ dr^2}{r^2f(r)}\,, \\
f(r) &=& 1- \frac{\pi^4 T^4}{r^4}\,,
\end{eqnarray}
where for simplicity of notation we have rescaled $t$ and $x$. This metric as a non degenerate horizon at
$r=r_H=\pi T$ with planar topology. The Hawking temperature is $T$. A natural conjecture is that this metric is now
the dual to the strongly coupled field theory in a state of thermal equilibrium at temperature $T$. More precisely it
corresponds to the plasma phase of the dual gauge theory \cite{,Gubser:1996de,Witten:1998zw}. For a conformal theory in Minkowski
space that any temperature puts the theory in the plasma phase. Because of conformal symmetry every finite temperature is as good as any other.
This is reflected in the gravity background by the fact that we can rescale the coordinates 
$r,t,\vec x$ and to set $T=1$. 

Suppose now that $\Phi$ is a small fluctuation around the black hole background. We expand it in plane boundary waves
\begin{equation}
 \Phi(r,t,\vec{x}) = \int \frac{d\omega dk}{(2\pi)^4} \tilde\Phi_0(\omega, \vec k ) e^{-i\omega t + i \vec{k}.\vec{x} }\, F_{\omega,k}(r)\,.
\end{equation}
For every fixed $\omega, \vec{k}$ the linearized equation of motion for the fluctuation boils down then to an ordinary
second order differential equation for $F_{\omega,\vec{k}}$. In turns out that in asymptotically (non-extremal) AdS black hole backgrounds this equation is of Fuchsian type, i.e. having at most a finite number of regular singular points in the complexified $r$-plane. 
The point at infinity is a regular singular point with characteristic exponents $\Delta_\pm$. 
Imposing infalling boundary conditions $ F_{\omega,k} \sim e^{-i\omega(t+r_*)}$ on the horizon (we use a tortoise coordinate
here $dr_* = dr/f(r)$ such that the horizon sits at $r_*\rightarrow -\infty$) the asymptotic expansion of a solution takes
the form
\begin{equation}
 F_{\omega,k} = A(\omega, \vec{k})\, r^{-\Delta_-} \left[1 + O(1/r)\right] + B(\omega,\vec k )\, r^{-\Delta_+} \left[ 1+ O(1/r)\right] \,.
\end{equation}
The retarded two point correlator turns out then to be given by the ratio of the two expansion coefficients $A,B$ \cite{Son:2002sd}
\begin{equation}\label{eq:Grholographic}
 G_R(\omega,\vec k ) = K \frac{B(\omega, \vec k )}{A(\omega, \vec k )} \,.
\end{equation}
This leads to the following holographic definition of quasinormal modes in AdS: they are solutions of the 
linearized fluctuation equations that correspond to poles of the holographic
retarded Green's function in the complexified frequency plane.

Retarded two point functions are the central objects in linear response theory. If we perturb a system by switching on
a small perturbation represented by a source $J(\omega,\vec k )$ that couples to the operator ${\cal O}$ the response
is given by 
\begin{equation}
 \left \langle {\cal O}(t,\vec{x})\right \rangle = \int \frac{d\omega d^3 k}{(2\pi)^4} \, e^{-i \omega t + i \vec{k}.\vec{x}} 
G_R(\omega,\vec k ) J(\omega, \vec k ) \,.
\end{equation}
If $G_R$ is a holographic Green's function computed from an asymptotically AdS black hole background with regular horizon 
it has indeed only poles in the complex frequency plane and the integral can be evaluated according to Cauchy's theorem 
\begin{equation}\label{eq:responseqnm}
 \left\langle {\cal O}(t,\vec{x}) \right \rangle = \Theta(t) \sum_n R_n(k) j(\omega_n(k)) e^{-i\omega_n t + i\vec{k}\vec{x} }\,,
\end{equation}
where $R_n$ are the residues of $G_R$ at the poles $\omega_n$. As long as all quasinormal frequencies lie in the lower half
of the complex $\omega$-plane the response is decaying exponentially fast. If one (or more) modes happen to lie in the upper half
plane we find a mode whose amplitude grows exponentially with time, getting soon out of the linearized regime and indicating an instability. 

\begin{figure}[!htpb]
 \includegraphics[height=7cm,width=10cm]{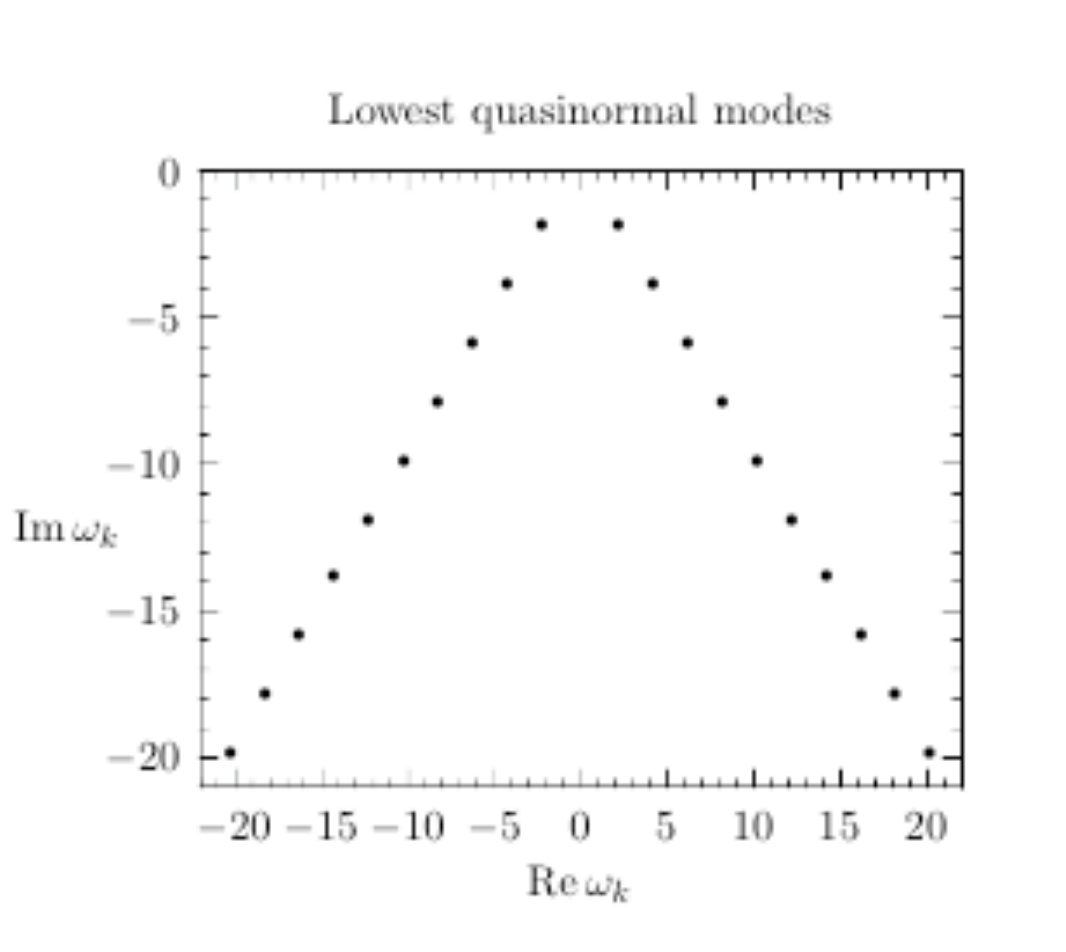}
\caption{The ten lowest lying quasinormal modes for a scalar field excitation (from \cite{Hoyos:2006gb}).  }
\end{figure}

If the operator $\cal O$ is a conserved charge then the spectrum of quasinormal modes contains special modes whose dispersion
relation fulfills 
\begin{equation}\label{eq:hydromode}
 \lim_{k\rightarrow 0} \omega_H(k) = 0\,.
\end{equation}
The reflect that fact that the total charge can not dissipate away. We call these special modes hydrodynamic.
These modes play a distinguished role in the applications of the AdS/CFT correspondence.

\begin{figure}[!htbp]
 \includegraphics[width=12cm,height=8cm]{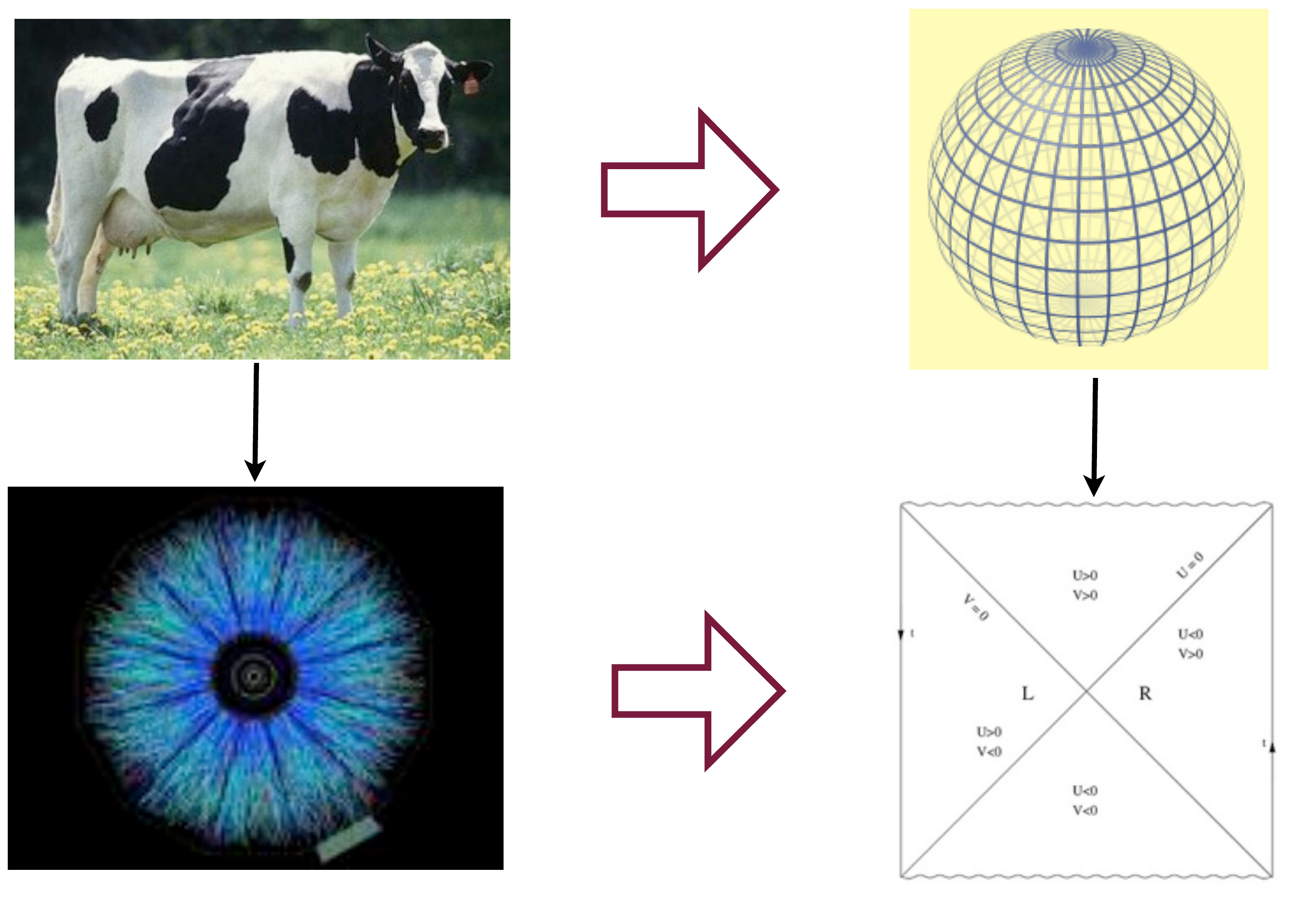}
\caption{Nature confronts us sometimes with complicated objects such as the one in the upper left
corner. In simple situations, as for example in collisions at ultrarelativistic energies, it is sufficient
to replace it by the simpler object in the right upper corner. The same philosophy can be applied
to the strongly coupled quark gluon plasma!}
\end{figure}

% \setlength{\unitlength}{3947sp}%
% %
% \begingroup\makeatletter\ifx\SetFigFont\undefined%
% \gdef\SetFigFont#1#2#3#4#5{%
%   \reset@font\fontsize{#1}{#2pt}%
%   \fontfamily{#3}\fontseries{#4}\fontshape{#5}%
%   \selectfont}%
% \fi\endgroup%
% \begin{picture}(8629,5538)(2161,-6166)
% \end{picture}%

\section{The holographic quark gluon plasma}

Let me come now to the first application of the ideas outlined in the previous section. What can quasinormal modes tell us about the physics of the quark gluon plasma? 

Strong nuclear interactions are on a fundamental level described by QCD, a non-abelian gauge theory with gauge group $SU(3)$ and
three quarks transforming in the fundamental representation. It is therefore a theory of the same type as the ${\cal N}=4$ supersymmetric gauge theory, however QCD is of course not conformal. Its gauge coupling runs leading to asymptotic freedom
in the UV and confinement in the IR. Its spectrum consists only composite color neutral objects such a Mesons and Baryons  at low energies. If
we are able however to increase the energy density to values such that $\sqrt[4]{\epsilon} \sim \Lambda_\mathrm{QCD}$ we expect
the theory to undergo a phase transition (or at least a fast crossover) to a deconfined regime. Such high energy densities can
be generated in experiments colliding heavy ions, i.e. gold or lead atoms completely strip off their electrons, at relativistic
energies. Two such experiments are currently running, a dedicated one at the Relativistic Heavy Ion Collider (RHIC) in the USA and
in Europe at the Large Hadron Collider (LHC) which every year dedicates its november run to heavy ion collision.

One of the spectacular findings of these experiments is that the droplet of hot matter created in the collision behaves as
an almost ideal fluid. It viscosity to entropy ratio seems to be the smallest one ever observed for a fluid. The value is
nowadays most often cited in units of $\frac{1}{4\pi}$, e.g. in \cite{Heinz:2011kt} we find
\begin{equation}
 1< 4\pi \frac{\eta}{s} < 2.5 \,. 
\end{equation}
These values are obtained by assuming that the quark gluon plasma can be described by hydrodynamics. Simulations based
on relativistic hydrodynamics where $\eta/s$ is an input parameter can be used to compute the particle multiplicities as
function of transverse momentum. Fitting these to the data allows the extraction of a value for $\eta/s$. This is of course
not the only input parameter, initial conditions play a role as do the time from when on the hydrodynamic description is
supposed to be valid. The latter is often referred to a thermalization time scale. Data can be well fitted assuming thermalization
times $0.6 fm/c < \tau < 1 fm/c$ \cite{Heinz:2011kt}. 

The temperatures reached in these heavy ion collisions are about two to three times the deconfinement temperature. This is 
a rather low value and indicates that QCD is still a rather strongly coupled system at these scales. Perturbative estimates
of $\eta/s$ and the thermalization time are in fact to high to be compatible with experimental data. We seem to be in need 
of a genuine strong coupling description of the quark gluon plasma. Since numerical methods based on lattice QCD are mostly
confined to Euclidean signature they are of limited help in calculating transport coefficients such as the viscosity.

We do have however an accessible model for a strongly coupled non-abelian gauge theory in the plasma phase via the AdS/CFT correspondence. It is of course quite different from QCD in its details but at least it might serve as a toy model. Let us see 
if we can subject this toy model to some simple tests that probe its applicability to the quark gluon plasma created in
heavy ion collisions. 

\subsection{Hydrodynamics}
Let us briefly recall the basics of hydrodynamics. It can be understood of an effective field
theory. The equations of motion are simply the conservation laws of energy-momentum and possible some conserved currents.
The energy-momentum tensor is expanded in the number of derivatives that act on the thermodynamic variables such as
pressure $p$, energy density $\epsilon$ and the fluid four velocity $u^\mu$. It is described by the equations
\begin{eqnarray}\label{eq:hydro}
 0 &=&\partial_\mu T^{\mu\nu} \,,\\
T^{\mu\nu} &=& (\epsilon + p) u^\mu u^\nu + p \eta^{\mu\nu} + \tau^{\mu\nu}\,,\\
\tau^{\mu\nu} &=& -\eta \sigma^{\mu\nu} - \zeta P^{\mu\nu} \partial_\rho u^\rho \,,\\
\end{eqnarray}
with the transverse projector $P^{\mu\nu} = u^\mu u^\nu + \eta^{\mu\nu}$ and the shear tensor $\sigma^{\mu\nu} = P^{\mu\alpha}
P^{\nu\beta} ( \partial_\alpha u_\beta + \partial_\beta u_\alpha - \frac 2 3 \eta_{\alpha\beta} \partial_\rho u^\rho)$.
Assuming the validity of the equation of state $p=p(\epsilon)$ at each space time point, i.e. local thermal equilibrium, 
we have four dynamical equations for the four variables $p$ and the normalized $u^\mu$. 
At this level there are two transport coefficients describing dissipation, the shear viscosity $\eta$ and the bulk viscosity
$\zeta$. It is not too difficult to see
that they have two eigenmodes with dispersion relations
\begin{eqnarray}
 \omega = -i \frac{\eta}{\epsilon + p} k^2 \label{eq:shearmode} \,,\\
 \omega = \pm v_s k - \Gamma_s k^2 \label{eq:soundmode}\,.
\end{eqnarray}
They are the shear mode and the sound mode, $v_s^2 = \frac{\partial p}{\partial \epsilon}$ is the speed of sound and
$\Gamma_s = \frac{\zeta + 4/3\eta}{2(\epsilon + p)}$ is the sound attenuation constant. Using the thermodynamic relation
$\epsilon+p = s T$ we see that $\eta/s$ determines the diffusion constant of the shear mode.

\subsection{Quasinormal modes}
Let me concentrate on the shear mode (\ref{eq:shearmode}). It obviously fulfills (\ref{eq:hydromode}). If the AdS black brane
is a strongly coupled plasma then we should be able to find such a mode in its quasinormal mode spectrum. This is indeed the
case. Since the source of the energy momentum tensor is the metric we switch on a metric perturbation $g_{MN}\rightarrow g_{MN} + h_{MN}$. We can chose the three momentum to lie in the $z$-direction. The relevant perturbation can then be packaged into a 
gauge invariant combination $Z = k_z h^t_x  + \omega h^z_x$ which in the coordinate $u=\frac{(\pi T)^2}{r^2}$ fulfills the
equation \cite{Kovtun:2005ev,Amado:2008ji}
\begin{equation}\label{eq:shearchannel}
 Z''(u) + \frac{w^2 - q^2 f(u))f(u) - u w^2 f'(u)}{u f(u) ( q^2 f(u) - w^2)} Z'(u) + \frac{w^2-q^2 f(u)}{u f(u)^2} Z(u) = 0\,,
\end{equation}
where $(w,q) = 2 \pi T (\omega, k)$ are dimensionless. In these coordinates quasinormal modes fulfill infalling boundary
conditions at the horizon at $u=1$ and Dirichlet boundary conditions $z=0$ at $u=0$.

\begin{figure}\label{fig:qnmshear}
 \includegraphics[height=10cm,width=10cm]{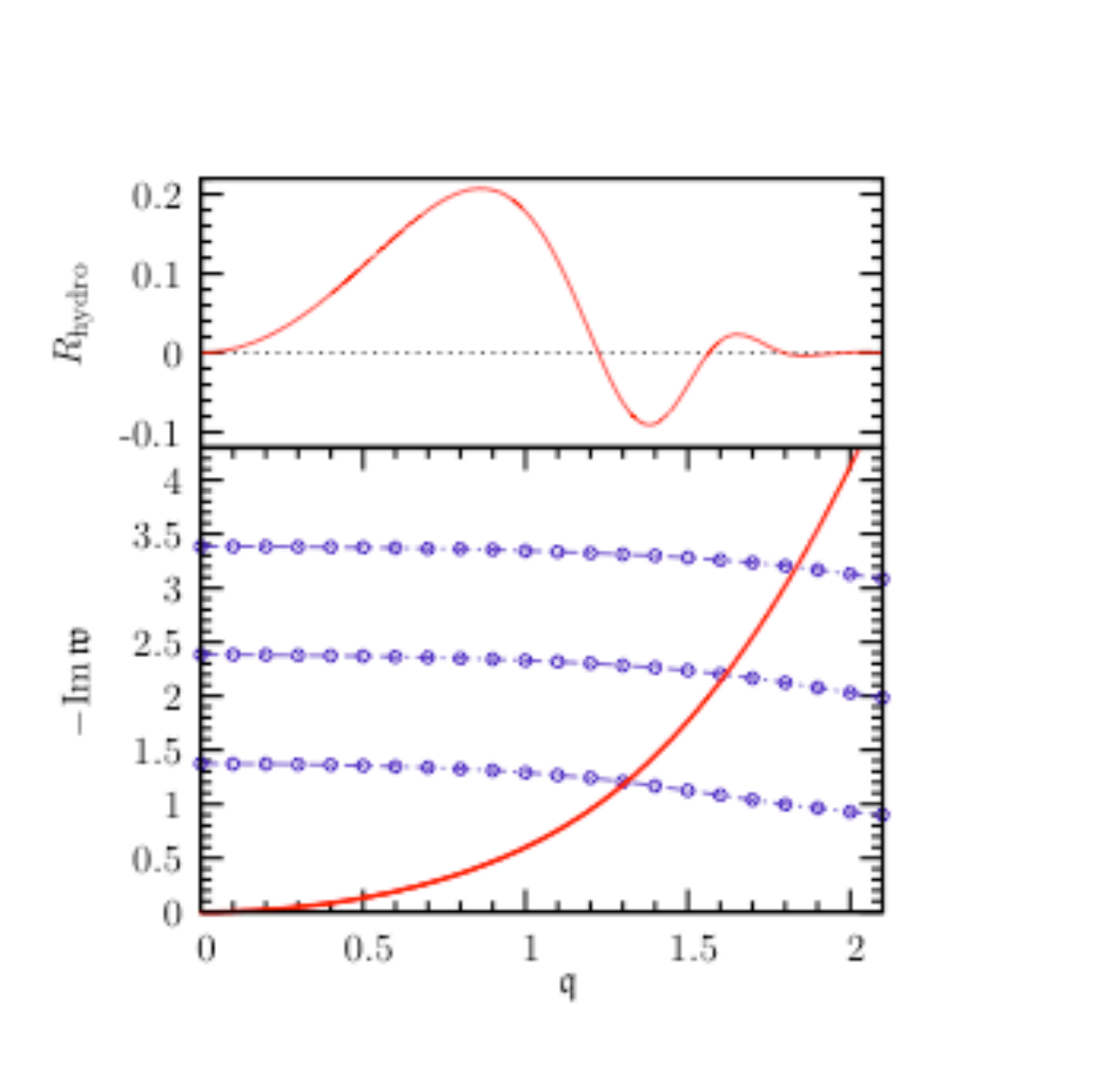}
\caption{Quasinormal modes in the shear channel as a function of momentum. In the lower part of the figure the imaginary
parts of the lowest lying modes are plotted. The ungapped mode shows a diffusion like dispersion relation at low momentum.
For at higher momentum or shorter wavelengths it crosses with the first gapped mode. The at these short wavelengths the system
does not behave according the hydrodynamics anymore. The upper part shows the residue of the diffusive mode. It goes to zero
as the momentum is increased.}
\end{figure}

{\bf Shear viscosity to entropy ratio:} The low lying spectrum is depicted in figure (\ref{fig:qnmshear}). It shows the imaginary parts of the four lowest lying quasinormal frequencies in the shear channel. There is one hydrodynamic one, whose frequency goes to zero with the momentum whereas
the others are gapped. The hydrodynamic mode is purely imaginary. The gapped modes also have non-vanishing real parts (not plotted). Hydrodynamics applies in the low frequency long wavelength limit. In this limit $\omega,k \rightarrow 0$ the equation
(\ref{eq:shearchannel}) can be solved analytically and one finds \cite{Policastro:2001yc, Policastro:2002se}
\begin{equation}
 \omega = -i \frac{1}{4\pi T} k^2\,.
\end{equation}
The shear viscosity to entropy ratio of the strongly coupled ${\cal N}=4$ plasma in the large $N$ limit is therefore
\begin{equation}
 \frac{\eta}{s} = \frac{1}{4\pi}\,.
\end{equation}
This we can take as a first success of our toy model! As mentioned this value is indeed compatible with the experimental
estimation and in any case much better than perturbative estimates. In fact not only the ${\cal N}=4$ theory gives this
value but many holographic models with less symmetries, e.g. less supersymmetry or non-conformal backgrounds give precisely
the same value. There is a certain amount of universality to this result. So far all holographic models with isotropic horizon and at most two derivative actions give this value \cite{Buchel:2003tz}. For some time it was conjectured that $1/(4\pi)$ is a fundamental lower bound \cite{Kovtun:2004de}.
If one includes however higher derivative terms the value can be even smaller \cite{Kats:2007mq,Brigante:2007nu} (see  \cite{Cremonini:2011iq} for a review of the effects higher derivative tems have on $\eta/s$). Very recently it has been shown that certain anisotropic backgrounds can violate the conjectured bound even at the two derivative level \cite{Rebhan:2011vd}. 

{\bf Validity of hydrodynamic approximation:} However from the figure we see that the system does not behave in a hydrodynamic way down to arbitrary small length scales.
This poses another test on our toy model. The length scale at which this happens should be much smaller than the size of
the fireball created in heavy ion collisions. Otherwise the toy model would predict that the fireball can not be
described by hydrodynamics! The fireball can be estimated to have a diameter of $10fm$. The temperature lies 
between $200$ MeV and $300$ MeV. The value at which the hydrodynamic mode crosses the first non-hydrodynamic mode
is $k=1.3 (2\pi T)$. We use the conversion factor $1$ fm$^{-1}$ = $197$ MeV and find a length scale of $l = (1.3 *2 *\pi * 200/197)^{-1} \approx 10^{-1}$ fm. This is good news! The fireball is approximately 100 times bigger than the length scale at which our strongly coupled toy model shows non-hydrodynamic behavior. Our toy model predicts therefore that hydrodynamic simulations should be
good approximations to capture the flow of the quark gluon plasma. 

{\bf Thermalization time scale:} In full generality this is of course a very complicated problem and relies on
a good modelling of the regime far from equilibrium beyond the linear response theory accessible to the quasinormal mode
analysis. Whatever this complicated far from equilibrium dynamics might be we can be sure however that at sufficiently late
times the system equilibrates in form of quasinormal mode excitations. Hydrodynamics is the late stage of this process
in which the lowest, hydrodynamic mode dominates the time development. We can estimate this time by comparing the 
first gapped with the hydrodynamic mode and demand that the response is dominated by the hydrodynamic mode from the time\footnote{This is sometimes also called the scrambling time.}
$\tau_H$ on. From (\ref{eq:responseqnm}) we can estimate 
\begin{equation}
 \tau_H = \frac{\log|R_H J(\omega_H)| - \log|R_1 J(\omega_1)|}{\Gamma_H - \Gamma_1}
\end{equation}
This depends of course on the particular form of the source $J$. By a judicial choice of source this time can be 
made as long as one wishes. For a variety of reasonable sources investigated in \cite{Amado:2007yr, Amado:2008ji} it turned out however 
that $\tau_H \approx 0.14$ fm/c. This again is good news. If this time had turned out to be much larger than $1$ fm/c
without having to adjust the sources in a special way it had meant that our toy model predicted generically thermalization
times larger than the values inferred from experiments.  

\begin{figure}\label{fig:frontvelocity}
 \includegraphics[height=6cm]{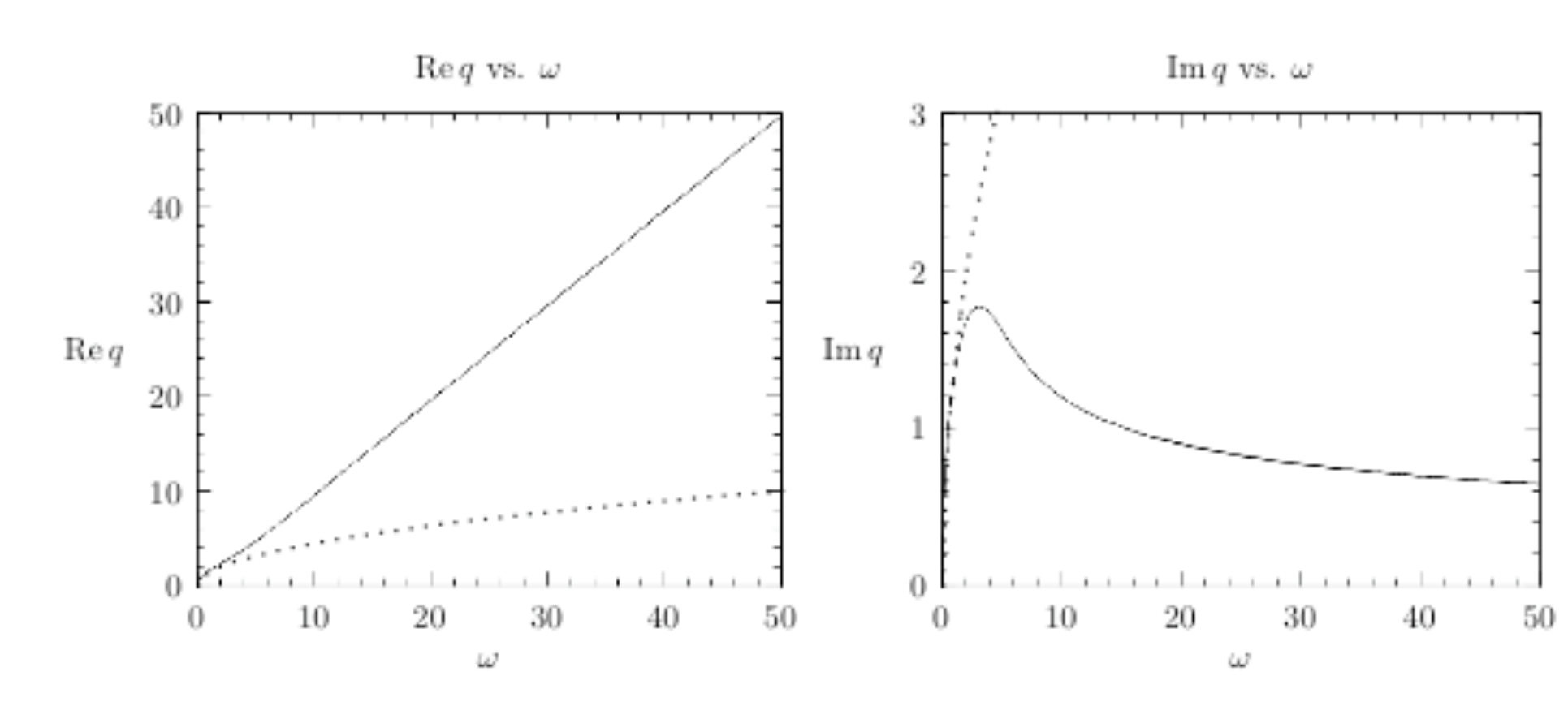}
\caption{Keeping the frequency real one can solve for complex momentum values. At small frequencies the behavior is
the typical diffusive one $\omega = -i D k^2$ whereas for large frequencies the front velocity $\lim_{\omega\rightarrow\infty} \frac{\omega}{\mathrm{Re}(k)}=1$  can be read off.} 
\end{figure}

I also want to comment on one more feature of the quasinormal mode spectrum depicted in the figure. If we go to large
momenta the prolongation of the hydrodynamic mode grows faster and faster. In fact it seems that the ratio $\omega/k$ 
grows as well and goes way beyond the causality bound of $1$. This looks a bit worrisome at first. But if we look at the 
residue of this mode we see that it goes to zero. So in the high momentum regime this mode effectively ceases to exist.
Furthermore to study the issue of causality one should really study the front velocity in which one keeps the frequency real
and searches for poles in the complex $k$ plane \cite{Amado:2007pv}. The large frequency behavior of
the mode hydrodynamic mode is perfectly causal with a front velocity $v_F=1$ (see figure (\ref{fig:frontvelocity}).

Let us summarize what we have seen from this simple considerations. The strongly coupled ${\cal N}=4$ gauge theory
seems to be a reasonable good toy model for the strongly coupled quark gluon plasma. 
It predicts that hydrodynamics is applicable to a quark gluon plasmaball as small as $10$ fm and the quasinormal modes
show decay rates that at least allow for a fast thermalization. It also predicts a shear viscosity
to entropy ratio compatible with the values inferred from data via hydrodynamic simulations. These are some of the most basic features of the holographic
model of the quark gluon plasma. There exist of course many more instances where quasinormal modes play a crucial role
in holographic models of the quark gluon plasma.
A comprehensive review on AdS/CFT applications to the physics of the strongly coupled quark gluon plasma can found in \cite{CasalderreySolana:2011us}.

%\subsection{Holographic meson melting}

\section{Holographic Superfluids}\label{sec:superfluid}

Superfluidity is one of the most emblematic phenomenon of many body physics. It is therefore quite natural to ask if
a holographic implementation of superfluidity can be found. On a fundamental level it is the question of spontaneous
symmetry breaking and it has been answered to the affirmative in \cite{Gubser:2008px, Hartnoll:2008kx}. A simple model
of a holographic superfluid is the one presented in \cite{Hartnoll:2008vx}. The model was first formulated in a $2+1$ dimensional
setting, i.e. in a four dimensional asymptotically AdS space. This is however not crucial and $3+1$ dimensional
models can easily be constructed. The quasinormal mode spectrum of this model has been investigated in \cite{Amado:2009ts} and
in the following we will briefly outline the construction of the model and the main features of its quasinormal mode
spectrum.  

One characteristic of superfluids is the phenomenon of second sound. Superfluid Helium II for example has the capacity
of transporting heat not in the usual diffusive form described by a heat equation but through propagating waves. This
is the phenomenon of second sound. We will review how this can be investigated in a holographic superfluid with the help
of quasinormal modes. 

We start with a four dimensional AdS black brane with metric
\begin{equation}
 ds^2 = - f(r) dt^2 + \frac{dr^2}{f(r)} + r^2\left( dx^2+dy^2\right)\,.
\end{equation}
where $f(r) = \frac{r^2}{L^2} - \frac{M}{r^2} $. This geometry has a regular horizon at $r_H= M^{1/3}L^{2/3}$ with
a Hawking temperature of $T= \frac{3}{4\pi} \frac{r_H}{L^2}$. 

In addition we will consider a simple Abelian Higgs model on this 
background with Lagrangian
\begin{equation}
 {\cal L} = -\frac{1}{4} F_{MN} F^{MN} - m^2 \Psi \bar \Psi - D_M\Psi D^M\Psi\,,
\end{equation}
$F_{MN}$ is the field strength of an abelian gauge field $A_M$ and $\Psi$ is a complex scalar field charged under
the $U(1)$ gauge symmetry. Its mass is chosen to be $m^2=-2L^2$ which would be tachyonic in flat space but is
perfectly stable in asymptotically AdS spaces. We will also consider only the so called decoupling limit in which
we neglect the back reaction of the gauge and scalar field onto the geometry. This can be justified in the limit
of large charge of the scalar field \cite{Hartnoll:2008vx}. 

Now one can look for non-trivial solutions to the field equations in which the temporal component of the
gauge field $A_0=\Phi$ and the scalar field
take the following form 
\begin{eqnarray}\label{eq:superasymptotics}
 \Phi &=& \frac{3 L}{4\pi T} \mu - \frac{9 L n}{16\pi^2T^2  }\frac{r_h^2}{r^2}+\cdots\,,\\
 \Psi &=& \psi_1\frac{ r_H }{r} + \psi_2 \frac{r_H^2}{ r^2 } + \cdots\,.
\end{eqnarray}
According to the holographic dictionary $\mu$ is interpreted as a chemical potential, $n$ the charge density, $\psi_1$ as
the source for a dimension $2$  operator $\psi_2= \frac{9}{16\pi^2 T^2 L^4} \langle {\cal O}_2\rangle$.

The underlaying conformal symmetry allows to fix one of the dimensionful parameters, we can chose e.g. $\mu=1$ and view
the variation of the dimensionless combination $\bar \mu = \frac{3 L}{4\pi} \frac{\mu}{T}$ as variations in the temperature.

It turns out that for high temperatures there is a unique solution with trivial scalar field $\Psi=0$ and gauge field
$\Phi = \bar \mu \left(1- \frac{ r_h}{r}\right)$. 

Below a certain critical temperature $T_c = 0.587 \mu$ there is however a second solution with non-trivial scalar field
profile with $\psi_1=0$ but $\psi_2\neq 0$. This is the sign of spontaneous symmetry breaking: although there is no source
for the operator ${\cal O}_2$ it does have a non trivial vacuum expectation value. 

\begin{figure}[!htbp]
\centering 
\begin{tabular}{cc}
\includegraphics[width=7cm]{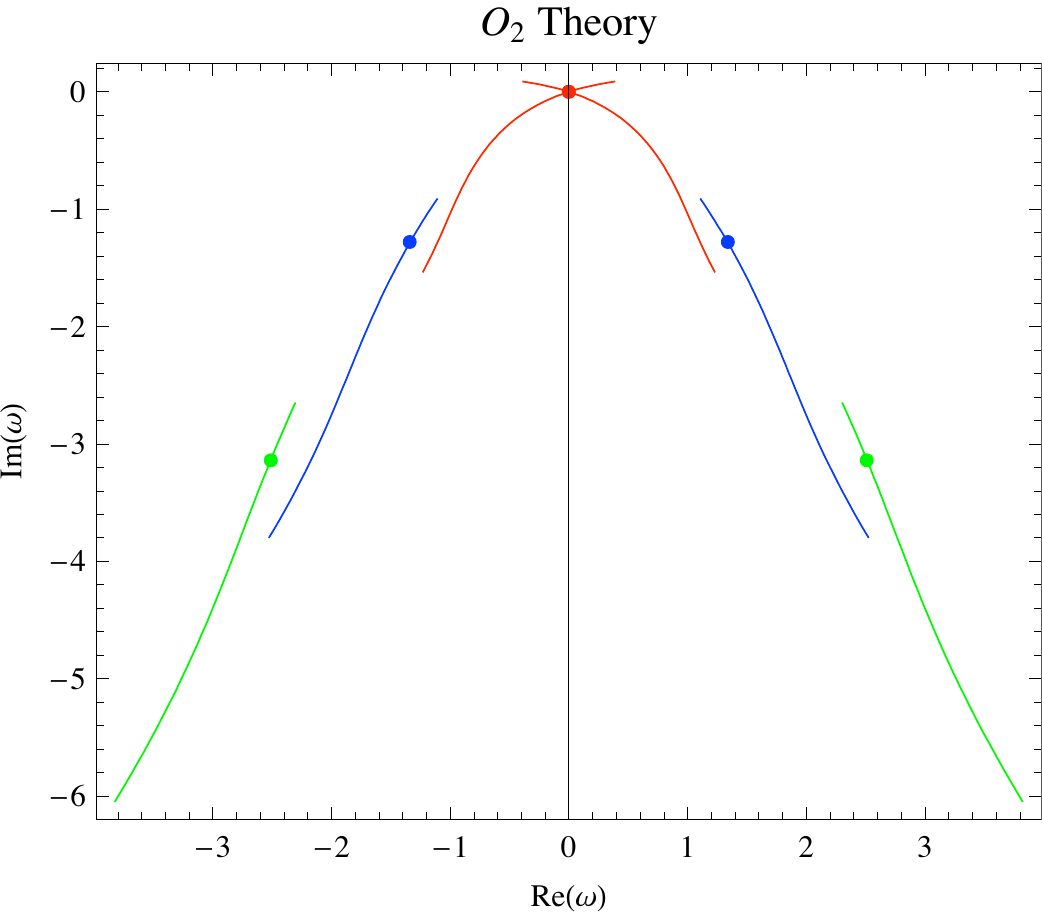}&
\includegraphics[width=7cm]{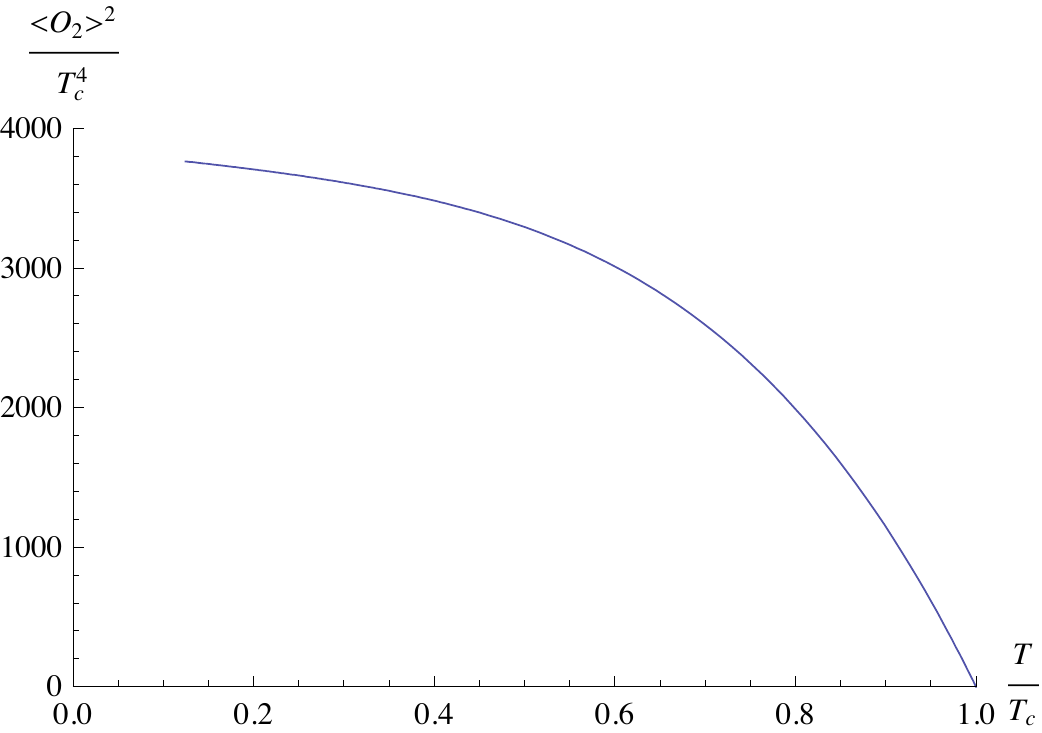} 
\end{tabular}
\caption{The figure on the left shows how the lowest quasinormal modes in the scalar field channel move as the 
temperature is varied. At the critical temperature the lowest two modes sit precisely at the origin, There they contribute
to the hydrodynamics. For even lower temperatures the move into the upper half plane indicating an instability towards condensation
the the scalar operator. The right figure shows the condensate as a function of Temperature. In this broken phase there are
no unstable modes.}
\end{figure}

In the normal phase the the quasinormal mode spectrum decomposes into three non-interacting channels: the scalar field channel,
the transverse gauge field and the longitudinal gauge field channel. The transverse gauge field channel shows a gapped spectrum,
the longitudinal one features one ungapped hydrodynamic mode. It reflects charge diffusion and has a diffusive behavior
of $\omega = -i D k^2$ with diffusion constant $D=\frac{3}{4\pi T}$. The most interesting behavior can be found in the 
scalar field channel. For high temperatures the spectrum is gapped but the gap diminishes as the temperature is lowered.
At the critical temperature two modes becomes ungapped, one for the scalar field $\Psi$ and one from the complex conjugate field
$\bar\Psi$. If one lowers the temperature even further but stays in the normal phase with vanishing expectation value for
the charged scalar operator these modes move into the upper half of the complex frequency plane. They are therefore exponentially
growing modes that signal an instability towards the formation of a scalar condensate! 

In the broken phase with non trivial $\Psi$ background the scalar field channel and the longitudinal vector field channel mix. Dividing the scalar in its real $\sigma$ and imaginary $\eta$ part the equations of motion become
{\small
\begin{eqnarray}\label{eq:coupledEOMsBroken}
0 &=&  f \eta''
 +\left(f'+\frac{2f}{\rho}\right) \eta'+\left(\frac{\phi^2}{f}+\frac{2}{L^2}
 +\frac{{ \omega}^2}{f}-\frac{{ k}^2}{\rho^2}\right) \eta 
 -\frac{2 i  \omega \phi }{f}\sigma
 -\frac{i\omega \Psi}{f}a_t-\frac{i  k \Psi}{r^2}a_x  \, , \nonumber \\
 &&\\
0 &=&  f \sigma''
 +\left(f'+\frac{2f}{\rho}\right) \sigma'+\left(\frac{\phi^2}{f}+\frac{2}{L^2}
 +\frac{{ \omega}^2}{f}-\frac{{ k}^2}{\rho^2}\right) \sigma 
 +\frac{2  \phi\Psi }{f} a_t+\frac{2 i  \omega \phi }{f}\eta \, , \\
0 &=& f {a_t}''+\frac{2f}{\rho}{a_t}'-\left(\frac{{ k}^2}{\rho^2}+2\Psi^2\right){a_t}
-\frac{ \omega k}{\rho^2}a_x-2 i  \omega\Psi\,\eta-4\Psi\phi\,\sigma  \, , \\
0 &=& f {a_x}'' +f'{a_x}'+\left(\frac{{ \omega}^2}{f}-2\Psi^2\right){a_x}
+\frac{ \omega  k}{f}a_t +2 i  k\Psi\, \eta\, . 
\end{eqnarray}
}
subject to the constraint $\frac{ \omega}{f}{a_t}'+\frac{k}{\rho^2}{a_x}'=2i\left(\psi'\,\eta-\psi\,\eta'\right)$.
One can construct three solutions obeying infalling boundary conditions at the horizon and fulfilling the constraint. 
A fourth one can be obtained by taking a pure gauge field configuration. These span a basis of the space of solutions
to (\ref{eq:coupledEOMsBroken}) which can be assembled in a $4\times 4$ matrix. The quasinormal mode conditions is
now that the determinant of this matrix evaluated at the boundary of the asymptotically AdS space vanishes. 

The goldstone theorem suggests that now there should be a massless mode in the spectrum with dispersion relation
\begin{equation}\label{eq:secondsound}
 \omega = \pm v_s k - i\Gamma_s k^2 \,,
\end{equation}
$v_s$ is the speed of second sound and $\Gamma_s$ is its attenuation constant. Careful search of the determinant condition
shows indeed that this mode is in the spectrum and that the purely imaginary mode that represented diffusion in the
unbroken phase is now gapped 
\begin{equation}
 \omega = -i \gamma -i \hat D k^2\,.
\end{equation}
So the only hydrodynamic modes in the broken phase are the Goldstone modes corresponding to second sound. The square
of the speed of sound and the gap in the purely imaginary mode close to the critical temperature behave as
\begin{equation}
 v_s^2\propto \left(1 - \frac{T}{T_c}\right) ~~~,~~~ \gamma\propto \left(1 - \frac{T}{T_c}\right)
\end{equation}

\begin{figure}[!htbp]
\centering
\begin{tabular}{cc}
 \includegraphics[width=7cm]{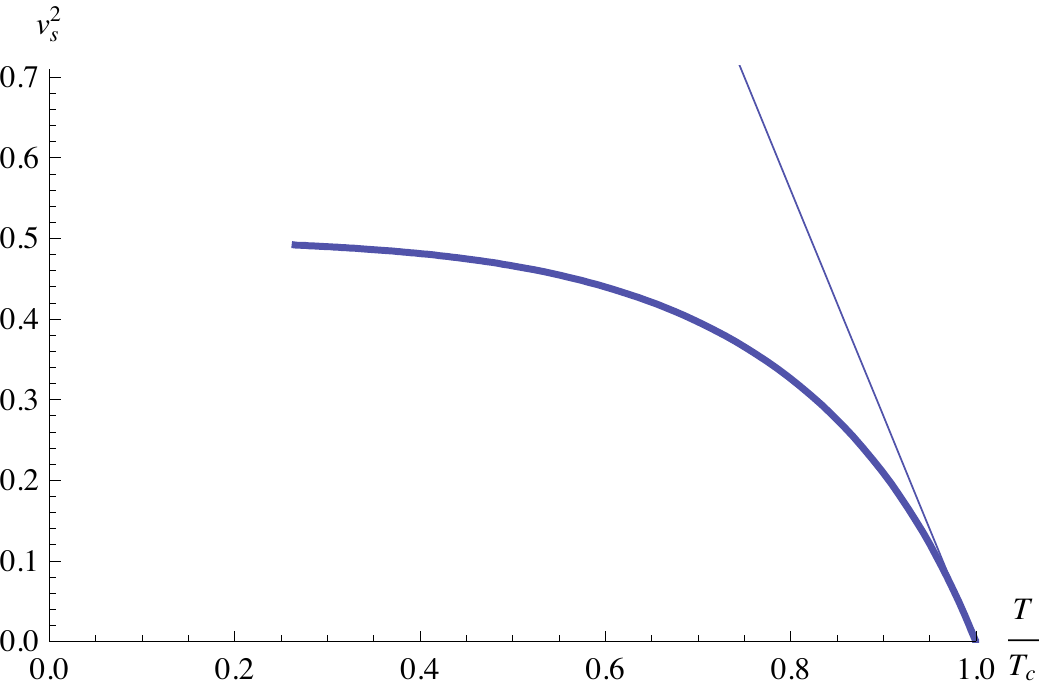} & \includegraphics[width=7cm]{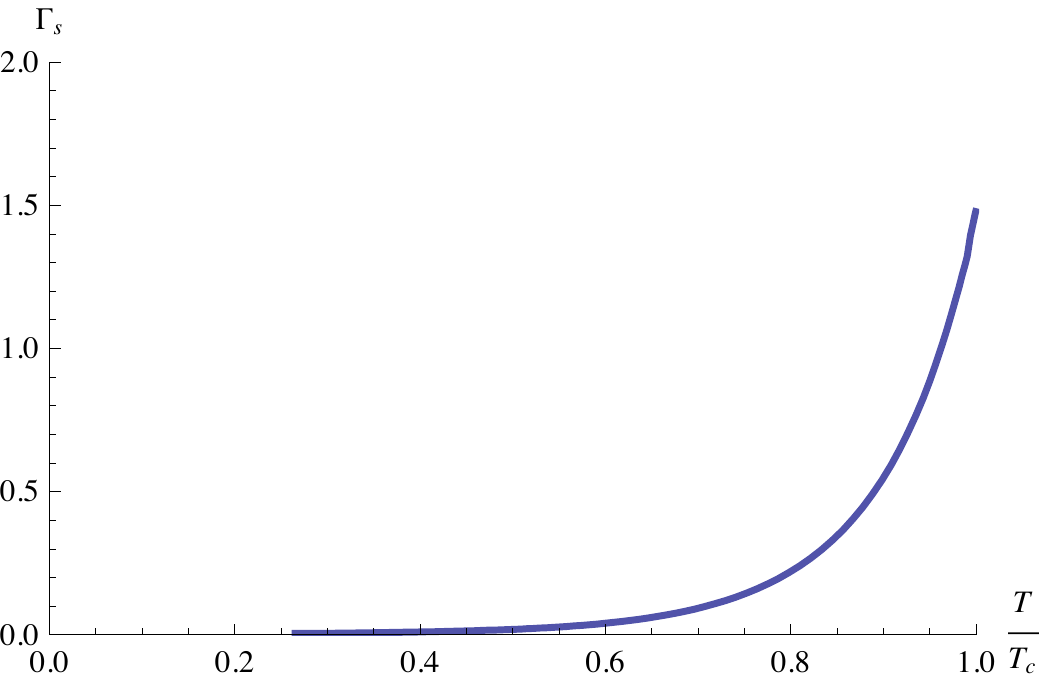}
\end{tabular}
\caption{The square of the speed of sound and its attenuation constant as determined from fitting the
lowed quasinormal mode to the dispersion relation (\ref{eq:secondsound}).\label{fig:secondsound}}
\end{figure}

The speed of sound and the attenuation constant can be obtained numerically by fitting the dispersion relation (\ref{eq:secondsound})
to the poles of the determinant as shown in figure (\ref{fig:secondsound}). As can be seen the attenuation constant rises
steeply close to the critical temperature but stays finite as one approaches it.

\begin{figure}[!htbp]
 \includegraphics[width=9cm]{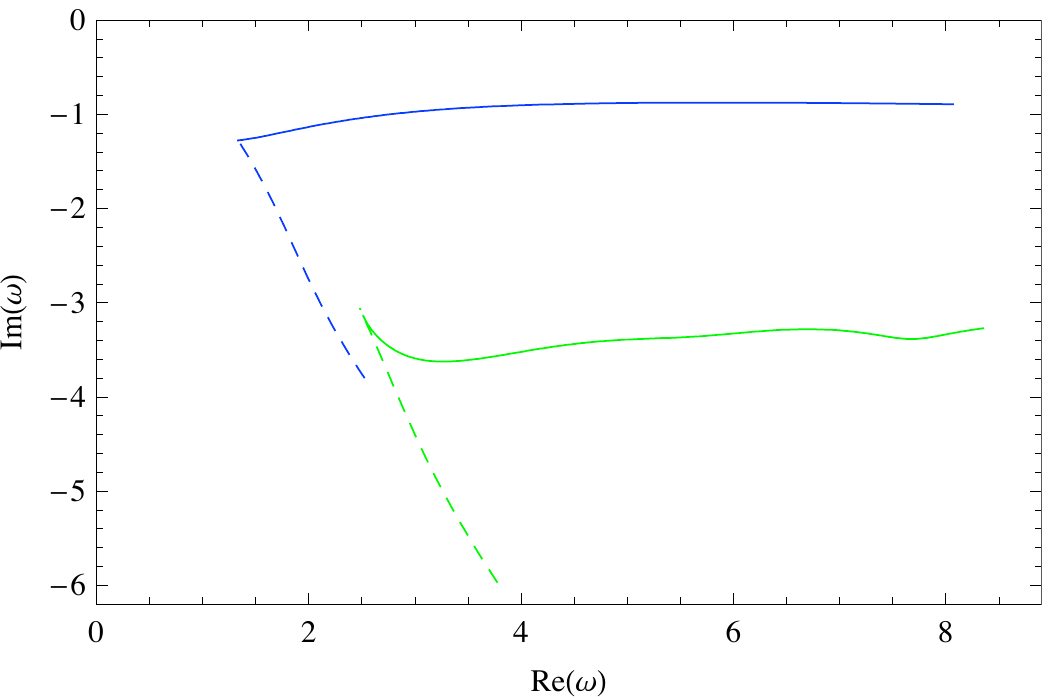} 
\caption{The higher, non-hydrodynamic modes move continuously but not smooth through the second order phase transition. 
\label{fig:highermodes}}
\end{figure}

The counting of hydrodynamic modes is as follows: in the normal phase there is one diffusive mode, precisely at
the critical temperature there are three modes, the diffusive one and the two scalarmodes that become ungapped
at the onset of the phase transition. In the broken phase the purely imaginary mode is gapped and there are two
ungapped modes corresponding to second sound.

Finally it is also of interest to see how the non-hydrodynamic modes behave as one goes through the phase transition.
this is depicted in figure (\ref{fig:highermodes}). The second and third scalar modes are shown as the critical temperature is
approached. They behave in a continuos but non-smooth way. For temperatures lower than the critical one the modes do not anymore
approach the real axis buy move roughly parallel to it as one lowers the temperature further. This suggests that no new instabilities arise. 

Applications of the AdS/CFT correspondence to condensed matter physics are reviewed in \cite{Hartnoll:2009sz,Herzog:2009xv, Horowitz:2010gk,McGreevy:2009xe, Iqbal:2011ae}

%Some url test \url{http://www.youtube.com/watch?v=xrL2ELkQOiE}.

%%%%%%%%%%%%%%%%%%%%%%%%%%%%%%%%%%%%%%%%%%%%
%% Sample figure:
%%
%% The option [height=...] scales the picture to the given height,
%% without it it would be printed at its nominal size
%%%%%%%%%%%%%%%%%%%%%%%%%%%%%%%%%%%%%%%%%%%%

% \begin{figure}
%   \includegraphics[height=.3\textheight]{golfer}
%   \caption{Picture to fixed height}
% \end{figure}

%%%%%%%%%%%%%%%%%%%%%%%%%%%%%%%%%%%%%%%%%%%%
%% SAMPLE TABLE
%%
%% Shows the use of \tablehead and \tablenote
%% macros
%%%%%%%%%%%%%%%%%%%%%%%%%%%%%%%%%%%%%%%%%%%%

% \begin{table}
% \begin{tabular}{lrrrr}
% \hline
%   & \tablehead{1}{r}{b}{Single\\outlet}
%   & \tablehead{1}{r}{b}{Small\tablenote{2-9 retail outlets}\\multiple}
%   & \tablehead{1}{r}{b}{Large\\multiple}
%   & \tablehead{1}{r}{b}{Total}   \\
% \hline
% 1982 & 98 & 129 & 620    & 847\\
% 1987 & 138 & 176 & 1000  & 1314\\
% 1991 & 173 & 248 & 1230  & 1651\\
% 1998\tablenote{predicted} & 200 & 300 & 1500  & 2000\\
% \hline
% \end{tabular}
% \caption{Average turnover per shop: by type
%   of retail organisation}
% \label{tab:a}
% \end{table}

%%%%%%%%%%%%%%%%%%%%%%%%%%%%%%%%%%%%%%%%%%%%%%%%
%% BACKMATTER
%%%%%%%%%%%%%%%%%%%%%%%%%%%%%%%%%%%%%%%%%%%%%%%%

\begin{theacknowledgments}
  I would like to thank I. Amado, J. Erdmenger, C. Greubel, C. Hoyos, M. Kaminski, P. Kerner, J. Mas, S. Montero, F. Pena-Benitez, J. Shock and  J. Tarrio for very enjoyable collaborations on the physics of quasinormal modes and the organizers of the ERE2011 conference for their kind invitation.
This work has been supported by Plan Nacional de Altas Energ\' ias FPA2009-07980, Consolider-Ingenio 2010 CPAN CSD2007-00042, HEP-HACOS S2009/ESP-2473. 

\end{theacknowledgments}



\bibliographystyle{aipproc}   % if natbib is available

\bibliography{ERE2011}

\hyphenation{Post-Script Sprin-ger}
\begin{thebibliography}{35}
\expandafter\ifx\csname natexlab\endcsname\relax\def\natexlab#1{#1}\fi
\providecommand{\enquote}[1]{``#1''}
\expandafter\ifx\csname url\endcsname\relax
  \def\url#1{\texttt{#1}}\fi
\expandafter\ifx\csname urlprefix\endcsname\relax\def\urlprefix{URL }\fi
\providecommand{\eprint}[2][]{\url{#2}}

\bibitem[Aharony et~al.(2000)]{Aharony:1999ti}
O.~Aharony, S.~S. Gubser, J.~M. Maldacena, H.~Ooguri, and Y.~Oz,
  \emph{Phys.Rept.} \textbf{323}, 183--386 (2000), \eprint{hep-th/9905111}.

\bibitem[Chan and Mann(1997)]{Chan:1996yk}
J.~Chan, and R.~B. Mann, \emph{Phys.Rev.} \textbf{D55}, 7546--7562 (1997),
  \eprint{gr-qc/9612026}.

\bibitem[Horowitz and Hubeny(2000)]{Horowitz:1999jd}
G.~T. Horowitz, and V.~E. Hubeny, \emph{Phys.Rev.} \textbf{D62}, 024027 (2000),
  \eprint{hep-th/9909056}.

\bibitem[Cardoso and Lemos(2001)]{Cardoso:2001hn}
V.~Cardoso, and J.~P. Lemos, \emph{Phys.Rev.} \textbf{D63}, 124015 (2001),
  latex, 14 pages, \eprint{gr-qc/0101052}.

\bibitem[Birmingham et~al.(2002)]{Birmingham:2001pj}
D.~Birmingham, I.~Sachs, and S.~N. Solodukhin, \emph{Phys.Rev.Lett.}
  \textbf{88}, 151301 (2002), \eprint{hep-th/0112055}.

\bibitem[Starinets(2002)]{Starinets:2002br}
A.~O. Starinets, \emph{Phys.Rev.} \textbf{D66}, 124013 (2002).

\bibitem[Nollert(1999)]{Nollert:1999ji}
H.-P. Nollert, \emph{Class.Quant.Grav.} \textbf{16}, R159--R216 (1999).

\bibitem[Berti et~al.(2009)]{Berti:2009kk}
E.~Berti, V.~Cardoso, and A.~O. Starinets, \emph{Class.Quant.Grav.}
  \textbf{26}, 163001 (2009), \eprint{0905.2975}.

\bibitem[Son and Starinets(2002)]{Son:2002sd}
D.~T. Son, and A.~O. Starinets, \emph{JHEP} \textbf{0209}, 042 (2002),
  \eprint{hep-th/0205051}.

\bibitem[Gubser et~al.(1996)]{Gubser:1996de}
S.~Gubser, I.~R. Klebanov, and A.~Peet, \emph{Phys.Rev.} \textbf{D54},
  3915--3919 (1996), \eprint{hep-th/9602135}.

\bibitem[Witten(1998)]{Witten:1998zw}
E.~Witten, \emph{Adv.Theor.Math.Phys.} \textbf{2}, 505--532 (1998),
  \eprint{hep-th/9803131}.

\bibitem[Hoyos-Badajoz et~al.(2007)]{Hoyos:2006gb}
C.~Hoyos-Badajoz, K.~Landsteiner, and S.~Montero, \emph{JHEP} \textbf{0704},
  031 (2007), \eprint{hep-th/0612169}.

\bibitem[Heinz et~al.(2011)]{Heinz:2011kt}
U.~W. Heinz, C.~Shen, and H.~Song  (2011), \eprint{1108.5323}.

\bibitem[Kovtun and Starinets(2005)]{Kovtun:2005ev}
P.~K. Kovtun, and A.~O. Starinets, \emph{Phys.Rev.} \textbf{D72}, 086009
  (2005), \eprint{hep-th/0506184}.

\bibitem[Amado et~al.(2008{\natexlab{a}})]{Amado:2008ji}
I.~Amado, C.~Hoyos-Badajoz, K.~Landsteiner, and S.~Montero, \emph{JHEP}
  \textbf{0807}, 133 (2008{\natexlab{a}}), \eprint{0805.2570}.

\bibitem[Policastro et~al.(2001)]{Policastro:2001yc}
G.~Policastro, D.~Son, and A.~Starinets, \emph{Phys.Rev.Lett.} \textbf{87},
  081601 (2001), \eprint{hep-th/0104066}.

\bibitem[Policastro et~al.(2002)]{Policastro:2002se}
G.~Policastro, D.~T. Son, and A.~O. Starinets, \emph{JHEP} \textbf{0209}, 043
  (2002), \eprint{hep-th/0205052}.

\bibitem[Buchel and Liu(2004)]{Buchel:2003tz}
A.~Buchel, and J.~T. Liu, \emph{Phys.Rev.Lett.} \textbf{93}, 090602 (2004),
  \eprint{hep-th/0311175}.

\bibitem[Kovtun et~al.(2005)]{Kovtun:2004de}
P.~Kovtun, D.~Son, and A.~Starinets, \emph{Phys.Rev.Lett.} \textbf{94}, 111601
  (2005), an Essay submitted to 2004 Gravity Research Foundation competition,
  \eprint{hep-th/0405231}.

\bibitem[Kats and Petrov(2009)]{Kats:2007mq}
Y.~Kats, and P.~Petrov, \emph{JHEP} \textbf{0901}, 044 (2009),
  \eprint{0712.0743}.

\bibitem[Brigante et~al.(2008)]{Brigante:2007nu}
M.~Brigante, H.~Liu, R.~C. Myers, S.~Shenker, and S.~Yaida, \emph{Phys.Rev.}
  \textbf{D77}, 126006 (2008), \eprint{0712.0805}.

\bibitem[Cremonini(2011)]{Cremonini:2011iq}
S.~Cremonini, \emph{Mod.Phys.Lett.} \textbf{B25}, 1867--1888 (2011),
  \eprint{1108.0677}.

\bibitem[Rebhan and Steineder(2012)]{Rebhan:2011vd}
A.~Rebhan, and D.~Steineder, \emph{Phys.Rev.Lett.} \textbf{108}, 021601 (2012),
  \eprint{1110.6825}.

\bibitem[Amado et~al.(2008{\natexlab{b}})]{Amado:2007yr}
I.~Amado, C.~Hoyos-Badajoz, K.~Landsteiner, and S.~Montero, \emph{Phys.Rev.}
  \textbf{D77}, 065004 (2008{\natexlab{b}}), \eprint{0710.4458}.

\bibitem[Amado et~al.(2007)]{Amado:2007pv}
I.~Amado, C.~Hoyos-Badajoz, K.~Landsteiner, and S.~Montero, \emph{JHEP}
  \textbf{0709}, 057 (2007), \eprint{0706.2750}.

\bibitem[Casalderrey-Solana et~al.(2011)]{CasalderreySolana:2011us}
J.~Casalderrey-Solana, H.~Liu, D.~Mateos, K.~Rajagopal, and U.~A. Wiedemann
  (2011), \eprint{1101.0618}.

\bibitem[Gubser(2008)]{Gubser:2008px}
S.~S. Gubser, \emph{Phys.Rev.} \textbf{D78}, 065034 (2008), \eprint{0801.2977}.

\bibitem[Hartnoll et~al.(2008{\natexlab{a}})]{Hartnoll:2008kx}
S.~A. Hartnoll, C.~P. Herzog, and G.~T. Horowitz, \emph{JHEP} \textbf{0812},
  015 (2008{\natexlab{a}}), \eprint{0810.1563}.

\bibitem[Hartnoll et~al.(2008{\natexlab{b}})]{Hartnoll:2008vx}
S.~A. Hartnoll, C.~P. Herzog, and G.~T. Horowitz, \emph{Phys.Rev.Lett.}
  \textbf{101}, 031601 (2008{\natexlab{b}}), \eprint{0803.3295}.

\bibitem[Amado et~al.(2009)]{Amado:2009ts}
I.~Amado, M.~Kaminski, and K.~Landsteiner, \emph{JHEP} \textbf{0905}, 021
  (2009), \eprint{0903.2209}.

\bibitem[Hartnoll(2009)]{Hartnoll:2009sz}
S.~A. Hartnoll, \emph{Class.Quant.Grav.} \textbf{26}, 224002 (2009),
  \eprint{0903.3246}.

\bibitem[Herzog(2009)]{Herzog:2009xv}
C.~P. Herzog, \emph{J.Phys.A} \textbf{A42}, 343001 (2009).

\bibitem[Horowitz(2010)]{Horowitz:2010gk}
G.~T. Horowitz  (2010), \eprint{1002.1722}.

\bibitem[McGreevy(2010)]{McGreevy:2009xe}
J.~McGreevy, \emph{Adv.High Energy Phys.} \textbf{2010}, 723105 (2010),
  \eprint{0909.0518}.

\bibitem[Iqbal et~al.(2011)]{Iqbal:2011ae}
N.~Iqbal, H.~Liu, and M.~Mezei  (2011), \eprint{1110.3814}.

\end{thebibliography}

\IfFileExists{\jobname.bbl}{}
 {\typeout{}
  \typeout{******************************************}
  \typeout{** Please run "bibtex \jobname" to optain}
  \typeout{** the bibliography and then re-run LaTeX}
  \typeout{** twice to fix the references!}
  \typeout{******************************************}
  \typeout{}
 }

\end{document}